\documentclass[aps,prl,showpacs,twocolumn,superscriptaddress]{revtex4}
\usepackage{graphicx}
\usepackage{xspace}
\usepackage{epsfig}
\usepackage{multirow}
\usepackage{dcolumn}  

\newcommand{\de}{\ensuremath{\Delta E}\xspace}

\newcommand{\bb}{\ensuremath{B \overline{B}}\xspace}

\def\myspecial#1{}                   
\def\calL{{\mathcal L}}

\def\Mbc{M_{\rm bc}}

\begin{document}

\myspecial{!userdict begin /bop-hook{gsave 300 50 translate 5 rotate
    /Times-Roman findfont 18 scalefont setfont
    0 0 moveto 0.70 setgray
    (\mySpecialText)
    show grestore}def end}




\title{\quad\\[0.5cm]
Measurements of Branching Fractions for $B \to K \pi$ and $B\to \pi 
\pi$ Decays with 449 million $B\overline{B}$ Pairs}

\affiliation{Budker Institute of Nuclear Physics, Novosibirsk}
\affiliation{Chiba University, Chiba}
\affiliation{University of Cincinnati, Cincinnati, Ohio 45221}
\affiliation{Department of Physics, Fu Jen Catholic University, Taipei}
\affiliation{The Graduate University for Advanced Studies, Hayama}
\affiliation{Gyeongsang National University, Chinju}
\affiliation{Hanyang University, Seoul}
\affiliation{University of Hawaii, Honolulu, Hawaii 96822}
\affiliation{High Energy Accelerator Research Organization (KEK), Tsukuba}
\affiliation{Institute of High Energy Physics, Chinese Academy of Sciences, Beijing}
\affiliation{Institute of High Energy Physics, Vienna}
\affiliation{Institute of High Energy Physics, Protvino}
\affiliation{Institute for Theoretical and Experimental Physics, Moscow}
\affiliation{J. Stefan Institute, Ljubljana}
\affiliation{Kanagawa University, Yokohama}
\affiliation{Korea University, Seoul}
\affiliation{Swiss Federal Institute of Technology of Lausanne, EPFL, Lausanne}
\affiliation{University of Ljubljana, Ljubljana}
\affiliation{University of Maribor, Maribor}
\affiliation{University of Melbourne, Victoria}
\affiliation{Nagoya University, Nagoya}
\affiliation{Nara Women's University, Nara}
\affiliation{National Central University, Chung-li}
\affiliation{National United University, Miao Li}
\affiliation{Department of Physics, National Taiwan University, Taipei}
\affiliation{H. Niewodniczanski Institute of Nuclear Physics, Krakow}
\affiliation{Nippon Dental University, Niigata}
\affiliation{Niigata University, Niigata}
\affiliation{University of Nova Gorica, Nova Gorica}
\affiliation{Osaka City University, Osaka}
\affiliation{Osaka University, Osaka}
\affiliation{Panjab University, Chandigarh}
\affiliation{Peking University, Beijing}
\affiliation{Princeton University, Princeton, New Jersey 08544}
\affiliation{RIKEN BNL Research Center, Upton, New York 11973}
\affiliation{Saga University, Saga}
\affiliation{University of Science and Technology of China, Hefei}
\affiliation{Seoul National University, Seoul}
\affiliation{Shinshu University, Nagano}
\affiliation{Sungkyunkwan University, Suwon}
\affiliation{University of Sydney, Sydney, New South Wales}
\affiliation{Toho University, Funabashi}
\affiliation{Tohoku Gakuin University, Tagajo}
\affiliation{Tohoku University, Sendai}
\affiliation{Department of Physics, University of Tokyo, Tokyo}
\affiliation{Tokyo Institute of Technology, Tokyo}
\affiliation{Tokyo Metropolitan University, Tokyo}
\affiliation{Tokyo University of Agriculture and Technology, Tokyo}
\affiliation{Virginia Polytechnic Institute and State University, Blacksburg, Virginia 24061}
\affiliation{Yonsei University, Seoul}
 \author{S.-W.~Lin}\affiliation{Department of Physics, National Taiwan University, Taipei} 
 \author{P.~Chang}\affiliation{Department of Physics, National Taiwan University, Taipei} 
 \author{K.~Abe}\affiliation{High Energy Accelerator Research Organization (KEK), Tsukuba} 
 \author{K.~Abe}\affiliation{Tohoku Gakuin University, Tagajo} 
 \author{I.~Adachi}\affiliation{High Energy Accelerator Research Organization (KEK), Tsukuba} 
 \author{H.~Aihara}\affiliation{Department of Physics, University of Tokyo, Tokyo} 
 \author{D.~Anipko}\affiliation{Budker Institute of Nuclear Physics, Novosibirsk} 
 \author{V.~Aulchenko}\affiliation{Budker Institute of Nuclear Physics, Novosibirsk} 
 \author{T.~Aushev}\affiliation{Swiss Federal Institute of Technology of Lausanne, EPFL, Lausanne}\affiliation{Institute for Theoretical and Experimental Physics, Moscow} 
 \author{S.~Bahinipati}\affiliation{University of Cincinnati, Cincinnati, Ohio 45221} 
% ...
\collaboration{The Belle Collaboration}

\noaffiliation

\begin{abstract}
We report measurements of branching fractions for $B \to K\pi$ and
$B\to \pi\pi$  decays based on a data sample of 449
million $\bb$ pairs collected at the $\Upsilon(4S)$ resonance with the Belle
detector at the KEKB asymmetric-energy $e^+ e^-$ collider.  
We also calculate the ratios of partial widths for the decays $B \to K\pi$,
namely $R_c = 1.08 \pm 0.06 \pm 0.08$ and $R_n = 1.08 \pm 0.08 ^{+0.09}_{-0.08}$
,where the first and the second errors are statistical and systematic,
respectively. These ratios are sensitive to enhanced electroweak
penguin contributions from new physics; the new measurements are, however,
consistent with Standard Model expectations. 
\end{abstract}

\maketitle

\tighten

{\renewcommand{\thefootnote}{\fnsymbol{footnote}} 
\setcounter{footnote}{0}

Tests of the Standard Model (SM) can be performed in $B$-meson decays
  to $K\pi$ and $\pi\pi$ final states, which involve various interplays
  between dominant $b\to u$ tree diagram, $b\to s$, $d$ penguin
  diagrams and other sub-dominant contributions.  In general, direct
  comparisons of the measured branching fractions with the SM
  predictions suffer from large hadronic uncertainties within the
  current theoretical framework.  However, many of the uncertainties
  cancel out in ratios of branching fractions. 
Previous experimental results \cite{bellebr,babarbr,cleobr} for the ratios 
$R_c\equiv 2\Gamma(B^+ \to K^+\pi^0)/\Gamma(B^+ \to K^0\pi^+)=1.00\pm 0.08$ and 
$R_n\equiv \Gamma(B^0 \to K^+\pi^-)/2\Gamma(B^0 \to K^0\pi^0) = 0.82\pm 0.08$ \cite{ratios}
deviate from the SM expectations within several 
approaches~\cite{buras,pqcd,rn1,rn2}. For example, Ref.~\cite{buras}
predicts the values $R_c = 1.15 \pm 0.05$ and $R_n = 1.12 \pm 0.05$, which are 
calculated 
assuming  $SU(3)$ flavor symmetry. If the differences between these SM
expectations and the measured values of $R_c$ and $R_n$ persist with more 
data, this would imply a large electroweak penguin contribution in
$B\to K\pi$ decays~\cite{buras,rn1,rn2}. 

In this letter, we report new measurements of the branching fractions for
$B \to K^+\pi^-$, $K^+ \pi^0$, $K^0 \pi^0$, $\pi^+ \pi^-$ and  $\pi^+ \pi^0$ 
decays with a data sample five times larger than that used in our previous 
study \cite{bellebr}. Recent Belle results for $B \to K\overline{K}$, $B^+ \to K^0 \pi^+$
and $B^0 \to \pi^0 \pi^0$ decays have been reported elsewhere \cite{kk,pi0pi0}. 
The results are based on a sample of (449.3 $\pm$ 5.7) $\times$ 10$^{6}$ $\bb$
pairs collected with the Belle detector at the KEKB $e^+e^-$ asymmetric-energy
(3.5 on 8~GeV) collider~\cite{kur}. 
The production rates of $B^+B^-$ and $B^0\overline{B}{}^0$ pairs are
assumed to be equal.
The inclusion of the charge-conjugate decay is implied, unless explicitly stated
otherwise.

The Belle detector is a large-solid-angle magnetic
spectrometer that consists of a silicon vertex detector (SVD),
a 50-layer central drift chamber (CDC), an array of
aerogel threshold Cherenkov counters (ACC),
a barrel-like arrangement of time-of-flight
scintillation counters, and an electromagnetic calorimeter
comprised of CsI(Tl) crystals located inside
a superconducting solenoid coil that provides a 1.5~T
magnetic field.  An iron flux-return located outside
the coil is instrumented to detect $K_L^0$ mesons and to identify
muons.  The detector is described in detail elsewhere~\cite{aba}.
Two different inner detector configurations were used. For the first sample
of 152 million $\bb$ pairs (set I), a 2.0 cm radius beampipe
and a three-layer silicon vertex detector were used;
for the latter  297 million $\bb$ pairs (set II),
a 1.5 cm radius beampipe, a four-layer silicon detector
and a small-cell inner drift chamber were used \cite{svd2}.

Primary charged tracks are required to have a distance of closest approach
to the interaction point (IP) of less than 4 cm in the beam
direction ($z$-axis) and less than 0.1 cm in  the transverse plane.
Charged kaons and pions are identified using $dE/dx$ information
from the CDC and Cherenkov light yields in the ACC, which are
combined to form a $K$-$\pi$ likelihood ratio $\mathcal{R}(K/\pi)
= \mathcal{L}_K/(\mathcal{L}_K+\mathcal{L}_\pi)$, where
$\mathcal{L}_{K}$ $(\mathcal{L}_{\pi})$ is the likelihood that the
track is a kaon (pion).
Charged tracks with $\mathcal{R}(K/\pi)>0.6$ ($<$0.4) are
 classified as kaons (pions). Typically, the kaon (pion) identification  
efficiency is 83\% (90\%), and 6\% (12\%) of selected kaons (pions) are
misidentified as pions (kaons).
Furthermore, we reject charged 
tracks that are consistent with an electron hypothesis. 
Candidate $K^0$ mesons are reconstructed as $K^0_S \to \pi^+ \pi^-$
decays with the branching fraction taken from Ref.~\cite{ksbr}.
We pair oppositely-charged tracks assuming the pion hypothesis
and require the invariant mass of the
pair to be within $\pm 18$ MeV/$c^2$ of the nominal $K_S^0$ mass. The
intersection point of the $\pi^+ \pi^-$ pair must be displaced from the IP \cite{ksdis}.
Pairs of photons with invariant masses in the range of 115 MeV/$c^2$ $<$ $M_{\gamma \gamma}$ $<$ 152 MeV/$c^2$ ($\pm 3 \sigma$) 
are considered as $\pi^0$ candidates. The photon energy
is required to be greater than 50 MeV in 
the barrel region, defined as $32^\circ <\theta_{\gamma}< 128^\circ$, and
greater than 100 MeV in the end-cap regions, defined as $17^\circ < 
\theta_{\gamma}< 32^\circ$ or $128^\circ <\theta_{\gamma} <150^\circ$,
where $\theta_{\gamma}$ denotes the photon polar angle with respect to the 
direction anti-parallel to the $e^+$ beam.

Candidate $B$ mesons are identified by the ``beam-energy-constrained'' mass,
$M_{\rm bc} \equiv
\sqrt{E^{*2}_{\mbox{\scriptsize beam}}/c^4 - p_B^{*2}/c^2}$, and the energy difference,
$\Delta E \equiv E_B^* - E^*_{\mbox{\scriptsize beam}}$, where
$E^*_{\mbox{\scriptsize beam}}$ is the run-dependent beam energy, and $E^*_B$ 
and $p^*_B$ are the reconstructed energy and momentum of the $B$ 
candidates in the center-of-mass (CM) frame, respectively. Events with
$M_{\rm bc} > 5.20$ GeV/$c^2$ and $|\Delta E| < 0.3~{\rm GeV}$
are selected for the analysis.

The dominant background is from $e^+e^- \to q\overline q ~( q=u,d,s,c )$ 
continuum events. We use event topology to distinguish the $B\overline{B}$
events from the jet-like continuum background. We combine a set of modified 
Fox-Wolfram moments \cite{pi0pi02} into a
Fisher discriminant. A signal/background likelihood is formed, based on a
GEANT-based~\cite{geant}
Monte Carlo (MC) simulation, from the product of the probability density
functions (PDFs) for the Fisher discriminant and that for the cosine of the 
polar angle of the $B$-meson flight direction.
Suppression of the
continuum is achieved by applying a requirement on the ratio
$\mathcal{R} = {\calL}_{\rm sig}/({\calL}_{\rm sig} + {\calL}_{q \overline{q}})$, where
${\mathcal L}_{\rm sig}$ (${\mathcal L}_{q \overline{q}}$)
is the signal (continuum) likelihood. 
Continuum background is further suppressed through use of the
$B$-flavor tagging algorithm \cite{tagging}, which provides a discrete
variable indicating the flavor of the tagging $B$ meson and a continuous 
quality parameter $r$ ranging from 0 (for no
flavor-tagging information) to 1 (for unambiguous flavor assignment).
Events with a high value of $r$ are considered well-tagged and hence are
unlikely to have originated from continuum processes.
We classify events separately as poorly-tagged ($r \leq 0.5$) and well-tagged
($r>0.5$) in
data set I and data set II and for each category we determine a continuum 
suppression requirement for $\cal R$ that maximizes the value of 
$N_{\rm sig}^{\rm exp}/\sqrt{N_{\rm sig}^{\rm exp}+N_{q \overline{q}}^{\rm exp}}$.
Here, $N_{\rm sig}^{\rm exp}$ denotes the expected signal yields based on MC 
simulation and the average branching fractions of the previous 
measurements \cite{bellebr,babarbr,cleobr}, and
$N_{q\overline{q}}^{\rm exp}$ denotes the
expected continuum yields as estimated from sideband data ($M_{\rm bc}<5.26$ GeV/$c^2$ and $|\Delta E|<$ 0.3 GeV).

Background contributions from $\Upsilon(4S) \to B\overline B$ events are
investigated using a large MC sample that includes events from $b\to c$
transitions and charmless $B$ decays. After all the selection requirements,
no $b\to c$ background is found, while a small contribution from 
charmless $B$ decays is present at low $\Delta E$ values for all studied modes. 
Due to $K-\pi$ misidentification, large $B^0 \to K^+ \pi^-$ and $B^+ \to K^+ \pi^0$  feed-across backgrounds appear in the $B^0 \to \pi^+ \pi^-$ and $B^+ \to 
\pi^+ \pi^0$ modes, respectively.

The signal yields are extracted by performing extended unbinned 
maximum likelihood fits to the ($M_{\rm bc}$, $\Delta E$)
distributions of the selected candidate events.
The likelihood function for each mode is defined as
\begin{eqnarray}
\mathcal{L} & = & \frac{{\rm exp}\; (-\sum_{l,k,j} N_{l,k,j})}{N!}
\prod_i (\sum_{l,k,j} N_{l,k,j} P^i_{l,k,j}),
\end{eqnarray}
where $N$ is the total number of events, $i$ is the event identifier, $l$ 
indicates set I or set II, $k$ distinguishes 
the two $r$ regions and $j$ runs over all components included in the fitting
function: signal, continuum background, feed-across, and charmless $B$ 
background.
The variable $N_{l,k,j}$ denotes the number of events, and $P^i_{l,k,j}= {\cal P}_{l,k,j}(M_{\rm bc}^i$, $\Delta E^i)$ are two-dimensional PDFs, which are the same in the two $r$ 
regions for all fit 
components except for the continuum background. 

All the signal PDFs ($P_{l,k,j=\mathrm{signal}}(M_{\rm bc},\Delta E)$)
are parameterized by smoothed two-dimensional histograms obtained from correctly
reconstructed signal
MC based on the set I and set II detector configurations.
Signal MC events are generated with the PHOTOS \cite{photos}
simulation package to take into account final-state radiation. 
Since the $M_{\rm bc}$ signal distribution is dominated by
the beam-energy spread, we use the signal-peak positions and resolutions
obtained from $B^+ \to \overline{D}{}^0\pi^+$ data to refine our signal MC
(the $\overline{D}{}^0 \to K^+ \pi^- \pi^0$ sub-decay is used for modes
with a $\pi^0$ in the final state, while
$\overline{D}{}^0 \to K^+\pi^-$ is used for the other modes). The resolution 
for the $\Delta E$ distribution is calibrated using the invariant 
mass distribution of high momentum ($p_{\rm Lab}>$ 3 GeV/$c$) $D$ mesons. 
The size of the final-state radiation effects can be assessed if we take
signal PDFs from MC without PHOTOS and use these PDFs to extract the signal
yields from the signal MC with PHOTOS. The extracted yields
decrease by 5.8\% for $B^0 \to K^+ \pi^-$, 9.4\% for $B^0 \to \pi^+ \pi^-$ and 3.6\% for $B^+ \to K^+ \pi^0$ and $B^+ \to \pi^+ \pi^0$,
respectively.

The continuum background PDF is described by a product of a linear function for 
$\Delta E$ and an ARGUS function, $f(x) = x \sqrt{1-x^2}\;{\rm exp}\;[ -\xi (1-x^2)]$, where
$x$ = $\Mbc c^2$/$E^*_{\rm beam}$ \cite{argus}.
The overall normalization, $\Delta E$ slope and ARGUS parameter $\xi$
are free parameters in the fit.
The background PDFs for charmless $B$ decays are modeled by a smoothed 
two-dimensional histogram, obtained from a large MC sample. 
We also use a smoothed two-dimensional histogram
to describe the feed-across background, since the background events have  
($\Mbc$, $\de$) shapes similar
to the signal, except for a $\de$ peak position shift of $\simeq45$ MeV.
We perform a simultaneous
fit for $B^0 \to K^+ \pi^-$ and $B^0 \to \pi^+ \pi^-$, since these two decay 
modes feed across into each other. The feed-across fractions are constrained 
according to the identification efficiencies and fake rates of kaons and pions.
A simultaneous fit is also used for the $B^+ \to K^+ \pi^0$
and $B^+ \to \pi^+ \pi^0$ decay modes.

\begin{figure}[!b]
\includegraphics[width=0.52\textwidth]{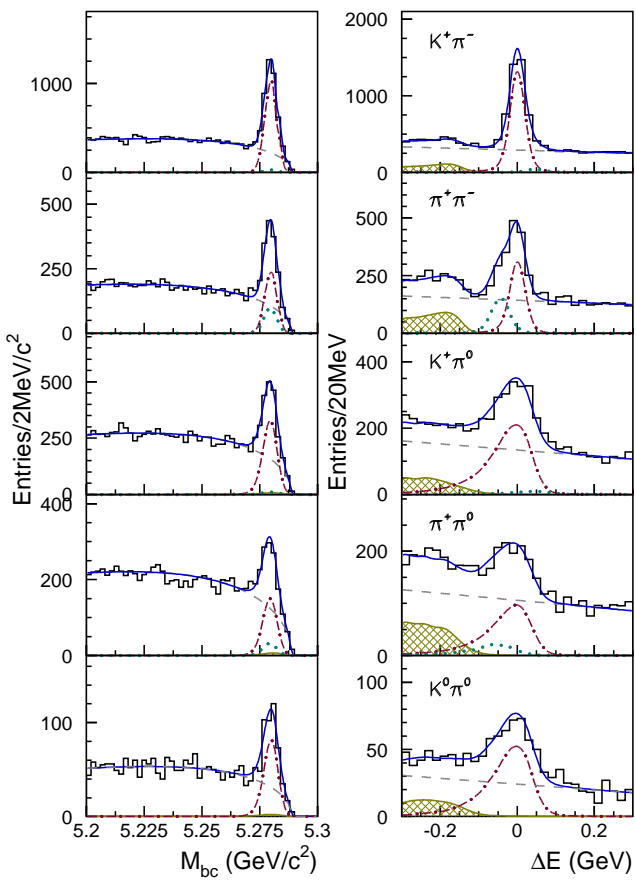}
\caption{$M_{\rm bc}$ (left) and $\Delta E$ (right) distributions for
$B^0\to K^+\pi^-$, $B^0 \to \pi^+ \pi^-$, $B^+ \to K^+ \pi^0$, $B^+ \to \pi^+ \pi^0$ and $B^0 \to K^0 \pi^0$ candidates. The histograms show 
the data, while the curves represent the various components from
the fit: signal (dot-dashed), continuum (dashed), charmless $B$ decays
(hatched), background from  mis-identification (dotted),
and sum of all components (solid). The $\Mbc$ and $\de$ projections of the fits
 are for events that have $|\de|< 0.06$ GeV (left) and 
$5.271$ GeV/$c^2 < {M_{\rm bc}} <5.289$ GeV/$c^2$ (right). (A looser 
requirement, $-0.14$ GeV $<\de<0.06$ GeV, is used for the modes with a $\pi^0$ 
meson in the final state.)}
\label{fig:kpi}
\end{figure}

When likelihood fits are performed, the yields
are allowed to
float independently for each $l$ (set I or set II) and $k$ bin (low
or high $r$ region). 
The $M_{\rm bc}$ and $\Delta E$ projections of the fits are shown in 
Fig.~\ref{fig:kpi}, while Table \ref{tab:br} summarizes the fit results for
each mode. The branching fraction of each mode is calculated by dividing the 
total signal yield by the number of $B\overline{B}$ pairs and by the average
reconstruction efficiency. The calculation of this average
efficiency takes into account the differences between various $l$ and
$k$ bins, and sub-decay branching fractions.

The fitting systematic errors are due to signal PDF modeling, 
charmless $B$ background modeling, and feed-across constraints. The first
 and last of these errors are estimated from the fit deviations after varying each 
parameter of the signal PDFs or the yields of the feed-across backgrounds by one
standard deviation. The effects due to fake-rate uncertainties are also 
included in the 
systematic error of the feed-across backgrounds. The systematic error due to the charmless $B$ background 
modeling is evaluated by requiring that $\Delta E > -0.12$ GeV, since the $\Delta E$ values of the charmless $B$ events are typically smaller than $-0.12$ GeV.
The above deviations in the signal yield are added in quadrature to obtain
the overall systematic error due to fitting. 

The MC-data efficiency difference due to the requirement on the likelihood 
ratio $\mathcal{R}$ is investigated with 
$B^+\to \overline{D}{}^0\pi^+ $ samples.
The systematic error due to the charged-track reconstruction efficiency is
estimated to be 1\% per track using partially
reconstructed $D^*$ events. 
The systematic error due to the $\mathcal{R}(K/\pi)$ selection is 1.3\% 
for pions and 1.5\% for kaons, respectively.
The $K_S^0$ reconstruction and the systematic error is verified by comparing the
ratio of $D^+\to K_S^0\pi^+$ and $D^+\to K^-\pi^+\pi^+$ yields
with the MC expectations. The $\pi^0$ reconstruction efficiency and the 
systematic error is verified by comparing the ratio of 
$\overline{D}{}^0 \to K^+ \pi^-$ and $\overline{D}{}^0 \to K^+ \pi^- \pi^0$
yields with the MC expectations.
Possible systematic uncertainties due to the 
description of final-state radiation have been studied by comparing the
latest theoretical calculations with the PHOTOS MC \cite{photoserr}. These 
uncertainties were found to be negligible and thus no systematic error is 
assigned due to PHOTOS. 
The systematic error due to the uncertainty of the total number of $\bb$ pairs 
is 1.3\% and the error due to signal MC statistics is between 0.4\%
and 0.7\%. 
The final systematic uncertainty is obtained by quadratically summing all
the contributions, as shown in Table~\ref{tab:sys}.

\begin{table}[h!]
\begin{center}
\caption{Extracted signal yields, product of efficiencies and sub-decay
branching ratios $({\cal B}_s)$, and calculated branching fractions for 
individual modes. The branching fraction errors are statistical and 
systematic, respectively.}
\begin{tabular}{lccccc}
\hline\hline
~Mode~ & Yield & Eff.$\times {\cal B}_s$(\%) & ${\cal B}(10^{-6})$ \\
\hline
~$K^+\pi^-$ &$ 3585^{+69}_{-68}$ & 40.16 &
$19.9 \pm 0.4 \pm 0.8$ \\
~$\pi^+ \pi^-$ & 872$^{+41}_{-40}$  & 37.98 & 5.1 $\pm 0.2 \pm 0.2 $ \\
~$K^+ \pi^0$ &$1493^{+57}_{-55}$ & $26.86$ &$12.4 \pm 0.5 \pm 0.6$  \\
~$\pi^+ \pi^0$ & 693$^{+46}_{-43}$ & 23.63 & 6.5 $\pm$ 0.4 $^{+0.4}_{-0.5}$  \\
~$K^0 \pi^0$ & 379$^{+28}_{-27}$ & 9.17 & 9.2 $\pm$ 0.7 $^{+0.6}_{-0.7}$  \\
\hline\hline
\end{tabular}
\label{tab:br}
\end{center}
\end{table}
\begin{table}[h!]
\begin{center}
\caption{Summary of systematic errors, given in percent.}
\begin{tabular}{lccccc}
\hline\hline
& $K^+ \pi^-$ & $\pi^+\pi^-$ & $K^+ \pi^0$ & $\pi^+ \pi^0$ & $K^0 \pi^0$ \cr
\hline
Signal PDF & $\pm$0.2 & $\pm$0.3 & $\pm$0.4 & $\pm$0.5 & $^{+0.3}_{-0.4}$\cr
Charmless $B$ background & $^{+0.0}_{-0.2}$ & $^{+0.6}_{-0.0}$ & $^{+0.0}_{-0.9}$ & $^{+0.0}_{-5.0}$ & $^{+0.0}_{-3.0}$ \cr
Feed-across background & $^{+0.4}_{-0.3}$ & $^{+2.2}_{-2.1}$ & $^{+0.7}_{-0.6}$ & $^{+2.5}_{-2.4}$ & 0.0 \cr
$\mathcal{R}$ requirement & $\pm$1.0 & $\pm$1.0 & $\pm$1.3 & $\pm$1.4 & $\pm$1.5 \cr
Tracking & $\pm$2.0 & $\pm$2.0 & $\pm$1.0 & $\pm$1.0 & 0.0\cr
$\mathcal{R}(K/\pi)$ requirement & $\pm$2.9 & $\pm$2.8 & $\pm$1.5 & $\pm$1.3 &  0.0 \cr
$K^0_S$ reconstruction & 0.0 & 0.0 & 0.0 & 0.0 & $\pm$4.9 \cr
$\pi^0$ reconstruction & 0.0 & 0.0 & $\pm$4.0 & $\pm$4.0 & $\pm$4.0\cr
\# of $B\overline{B}$ & $\pm$1.3 & $\pm$1.3 & $\pm$1.3 & $\pm$1.3 & $\pm$1.3 \cr
Signal MC statistics & $\pm$0.6 & $\pm$0.4 & $\pm$0.4 & $\pm$0.5 &  $\pm$0.7\cr
\hline
Total & $\pm$4.0 & $^{+4.5}_{-4.4}$ &  $^{+4.8}_{-4.9}$ & $^{+5.4}_{-7.3}$ & $^{+6.7}_{-7.3}$ \cr
\hline
\hline
\end{tabular}
\label{tab:sys}
\end{center}
\end{table}

\begin{table}[h!]
\begin{center}
\caption{Partial width ratios of $B \to K \pi$ and $\pi \pi$ decays. The errors
are quoted in the same manner as in Table \ref{tab:br}.  }
\begin{tabular}{lcccccccc}\hline \hline
Modes & Ratio \cr
\hline
2$\Gamma(K^+ \pi^0)$/$\Gamma(K^0\pi^+)$ & 1.08 $\pm$ 0.06 $\pm$ 0.08\cr
$\Gamma(K^+ \pi^-)$/2$\Gamma(K^0 \pi^0)$ & 1.08 $\pm$ 0.08 $^{+0.09}_{-0.08}$ \cr
$\Gamma(K^+ \pi^-)$/$\Gamma(K^0\pi^+)$ & 0.94 $\pm$ 0.04 $\pm$0.05 \cr
$\Gamma(\pi^+\pi^-)$/$\Gamma(K^+ \pi^-)$ & 0.26 $\pm$ 0.01 $\pm$ 0.01 \cr
$\Gamma(\pi^+\pi^-)$/2$\Gamma(\pi^+ \pi^0)$ & 0.42 $\pm$ 0.03 $^{+0.03}_{-0.02}$  \cr
$\Gamma(\pi^+ \pi^0)$/$\Gamma(K^0 \pi^0)$ & 0.66 $\pm$ 0.07 $\pm$ 0.05 \cr
2$\Gamma(\pi^+\pi^0)$/$\Gamma(K^0\pi^+)$ & 0.57 $\pm$ 0.04 $^{+0.04}_{-0.05}$  \cr
\hline
\hline
\end{tabular}
\label{ratio}
\end{center}
\end{table}

The ratios of partial widths can be used to extract the angle $\phi_3$ and to 
search for new physics~\cite{buras,rn1,rn2}.
These ratios (listed in Table \ref{ratio}) are obtained from the five measurements in 
Table \ref{tab:br}  and the new measurement of 
${\cal B}(B^+\to K^0\pi^+) = (22.8 ^{+0.8}_{-0.7} \pm 1.3)\times 10^{-6}$
described in Ref. \cite{kk}.
The ratio of charged to neutral $B$ meson lifetime, $\tau_{B^+}$/$\tau_{B^0}$ = 1.076 $\pm$ 0.008 \cite{ratios}, is used to convert the
branching-fraction ratios into the ratios of partial widths. The total errors are reduced because of the cancellation of some common systematic errors.
With a factor of five times more data than that used for our previous published
results \cite{bellebr}, the statistical errors on the branching fractions for 
all decay modes are reduced by more than a factor of 2.3.
The central value of the $K^0\pi^0$ branching fraction has decreased from 
$11.7\times 10^{-6}$ to $9.2\times 10^{-6}$ and the $K^+\pi^-$ branching 
fraction has increased from $18.5\times 10^{-6}$ to $19.9\times 10^{-6}$, 
resulting in a change in $R_n$ from $0.79 \pm 0.18$ to $1.08\pm 0.12$. The
obtained value of $R_c= 1.08\pm 0.09$ is similar to the previous Belle
measurement ($1.09\pm 0.19$) but is more precise. The errors for 
$R_n$ and $R_c$ shown here are the  sum in quadrature of the statistical  
and systematic errors. These two ratios  are now consistent with  SM 
expectations \cite{buras,pqcd,rn1,rn2}.

In conclusion, we have measured the branching fractions for $B \to K \pi$ 
and $B\to \pi\pi$ decays with 449 million $\bb$ 
pairs collected at the $\Upsilon(4S)$ resonance with the Belle detector.
We confirm the expected
hierarchy of branching fractions : $\cal{B}$$(K^0 \pi^+)$ $\ge$ $\cal{B}$$(K^+\pi^-)$ $>$ $\cal{B}$$(K^+ \pi^0)$ $\ge$ $\cal{B}$$(K^0\pi^0)$ $>$ $\cal{B}$$(\pi^+ \pi^0)$ $\ge$ $\cal{B}$$(\pi^+ \pi^-)$ and find no significant deviation from SM expectations in the ratios of partial widths.
We also find that the ratios $R_n$ and $R_c$ are both in good agreement with SM
expectations, in contrast to early measurements \cite{bellebr,babarbr,cleobr}.

We thank the KEKB group for excellent operation of the
accelerator, the KEK cryogenics group for efficient solenoid
operations, and the KEK computer group and
the NII for valuable computing and Super-SINET network
support.  We acknowledge support from MEXT and JSPS (Japan);
ARC and DEST (Australia); NSFC and KIP of CAS (China);
DST (India); MOEHRD, KOSEF and KRF (Korea);
KBN (Poland); MIST (Russia); ARRS (Slovenia); SNSF (Switzerland);
NSC and MOE (Taiwan); and DOE (USA).

\end{document}